# Nanoscale Engineering of Wurtzite Ferroelectrics: Unveiling Phase Transition and Ferroelectric Switching in ScAlN Nanowires

Ding Wang,[1*a)] Ping Wang,[1*] Shubham Mondal,[1] Mingtao Hu,[1] Yuanpeng Wu,[1] Danhao Wang,[1] Kai Sun,[2] Zetian Mi[1, a]

[1] *Department of Electrical Engineering and Computer Science, University of Michigan, Ann Arbor, MI 48109, USA*

[2] *Michigan Center for Materials Characterization (MC)², University of Michigan, Ann Arbor, MI 48109, USA*

*To whom correspondence should be addressed: dinwan@umich.edu, ztmi@umich.edu*

[*] *Those two contribute equally to this work.*

## Abstract:

The pursuit of extreme device miniaturization and the exploration of novel physical phenomena have spurred significant interest in crystallographic phase control and ferroelectric switching in reduced dimensions. Recently, wurtzite ferroelectrics have emerged as a new class of functional materials, offering intriguing piezoelectric and ferroelectric properties, CMOS compatibility, and seamless integration with mainstream semiconductor technology. However, the exploration of crystallographic phase and ferroelectric switching in reduced dimensions, especially in nanostructures, has remained a largely uncharted territory. In this study, we present the first comprehensive investigation into the crystallographic phase transition of ScAlN nanowires across the full Sc compositional range. While a gradual transition from wurtzite to cubic phase was observed with increasing Sc composition, we further demonstrated that a highly ordered wurtzite phase ScAlN could be confined at the ScAlN/GaN interface for Sc contents surpassing what is possible in conventional films, holding great potential to addressing the fundamental high coercive field of wurtzite ferroelectrics. In addition, we provide the first evidence of ferroelectric switching in ScAlN nanowires, a result that holds significant implications for future device miniaturization. Our demonstration of tunable ferroelectric ScAlN nanowires opens new possibilities for nanoscale, domain, alloy, strain, and quantum engineering of wurtzite ferroelectrics, representing a significant stride towards the development of next-generation, miniaturized devices based on wurtzite ferroelectrics.

The exploration of crystallographic phase transition behavior, especially in low-dimensional forms, has emerged as a fertile ground for the discovery of new physical properties and the development of innovative device concepts. [1-8] A key area of interest in this field is the study of ferroelectrics, which have shown considerable promise for the development of future power-efficient and highly integrated devices and systems. [9-11] A recent breakthrough discovery has shown that conventional III-nitride semiconductors can be transformed to be ferroelectric with widely tunable properties by alloying with rare-earth elements such as scandium, boron and yttrium. [12-15] Theoretical predictions suggest that ScAlN can maintain a wurtzite phase structure for Sc compositions up to 56%, with an energy bandgap that varies from ~6.2 eV to ~4 eV. [16] The piezoelectric coefficient, [17] permittivity, [18] and optical $\chi^{(2)}$ properties [19] can be dramatically enhanced with increasing Sc incorporation in ScAlN, accompanied by a significant reduction of the coercive field. [20] These tunable properties, combined with the use of standard epitaxial growth techniques [21-22], make ScAlN a unique platform for next-generation electronics, acousto-electronics, and quantum photonics. [23-31]

To date, however, achieving high-quality wurtzite ferroelectric semiconductors has remained a formidable challenge especially for high Sc contents wherein nonessential secondary phases take place. [16-17] To avoid the formation of the secondary rock-salt phase, various growth approaches, including N-rich growth condition, metal-modulated epitaxy, and Ga-assisted epitaxy have been commonly adopted. [32-33] Yet the achievement of reasonable quality wurtzite ScAlN has been often limited to Sc composition <40% either by sputtering or by epitaxial growth. [22, 34] Alternatively, the efficient strain relaxation and limited area epitaxy in low-dimensional nanostructures promotes the formation of dislocation-free, pristine heterostructures with tunable properties, as demonstrated recently for high efficiency nanowire green and red micro-LEDs. [35-36] Therefore, understanding the crystallographic phase transition in the material with different Sc compositions, especially in low-dimensional forms, is of great interest and importance to guide future material and device developments and to bracket the design margin of this new class of ferroelectrics. [37] Besides, while efforts have been taken to reducing the thickness of ScAlN with promising results down to ~5 nm, [38-41] the lateral scaling down and size limit of wurtzite ferroelectrics, which is pivotal for future power-efficient and highly-integrated device applications, remain an underexplored frontier. [42] We note that there have been previous reports on ScAlN based transistors with narrow channels or short gate lengths, which, however, do not necessarily reflect the size of the switchable domains. [27, 43]

In this research, we offer an inaugural, in-depth exploration of the crystallographic phase transitions in ScAlN nanowires throughout the entire Sc compositional range. A distinct, gradual shift from the wurtzite to the cubic phase is observed as the Sc composition increased. Our findings further reveal that a well-ordered wurtzite phase ScAlN can be maintained at the ScAlN/GaN interface, even when the Sc concentration (45-50%) far exceeds levels achievable in bulk layers (which is typically below 40%). This unique feature could be harnessed to confine high quality ferroelectric phase to breed novel material and device concepts. More importantly, we provide the first unambiguous proof of ferroelectric switching in ScAlN nanowires of nanometer dimensions, an achievement with profound ramifications for the aggressive scaling of future devices. The successful synthesis of pristine, tunable ScAlN nanowires heralds unique opportunities for the nanoscale, domain, alloy, and quantum control and engineering of ferroelectrics. This marks a substantial advancement towards the development of next-generation, nanoscale devices leveraging wurtzite ferroelectrics.

**Experiment section:**

All the nanowire samples were grown using a Veeco GENxplor molecular beam epitaxy (MBE) system, equipped with Knudsen effusion cells for aluminum (Al, purity 99.99995%), gallium (Ga, purity 99.99999%), and scandium (Sc, purity 99.999%) sources, and a radio frequency (RF) plasma cell for nitrogen ($N_2$, purity 99.9999%) source. In this study, the nanowire structures were grown directly on n-type Si wafers. The surface oxidation layer was removed by etching the Si wafers in buffered HF for 2 min prior to loading into the MBE system. Subsequently, the substrates were baked and outgassed at 200 °C and 600 °C for 2 h in the MBE load-lock and preparation chamber, respectively. After loading into the growth chamber, the Si substrates were further heated up to 900 °C to remove the native oxide. The growth temperature was calibrated by monitoring the (7 × 7) to (1 × 1) reconstruction transition of Si(111) at ~830 °C using in-situ reflection high-energy electron diffraction (RHEED) system. Self-assembled GaN nanowires with a height of about 400 nm was first grown at 750 °C as the epitaxy template, [44] followed by the growth of ScAlN with a nominal thickness of about 200 nm. The Ga beam equivalent pressure (BEP) was maintained at $1.0 \times 10^{-7}$

Torr for the GaN nanowire growth, while the Sc (Al) BEP was varied from 0 to $3.0 \times 10^{-8}$ Torr ($3.0 \times 10^{-8}$ to 0 Torr) to control the Sc content (from 0 to 1). The GaN layer was doped with Si at a dopant cell temperature of 1200 °C to serve as the bottom electrode for conducting ferroelectric characterization. Energy dispersive X-ray spectroscopy (EDS) was used to calibrate the Sc content in ScAlN in a Hitachi SU8000 scanning electron microscope (SEM). Scanning transmission electron microscopy (STEM) specimens were prepared by a Thermo Fisher Scientific Helios G4 UXe focus ion beam (FIB) and STEM images were collected using a JEOL 3100R05 aberration-corrected S/TEM operated at 300 kV, with a collection range of 59-200 mrad for high-angle annular dark-field (HAADF) imaging. Ti/Al electrodes were lithographically patterned after opening circular windows with diameters of 3-50 μm on 200-nm-thick $SiO_2$ deposited on the nanowire surface using a GSI ULTRADEP 2000 Plasma Enhanced Chemical Vapor Deposition (PECVD) system. The P-V loops were collected using a Radiant Precision Premier II ferroelectric tester driven from the top electrode, while the piezo-response force microscopy (PFM) was done using Veeco Dimension Icon Atomic Force Microscope. During measurements the tip frequency was 15 kHz and $V_{AC}$ was 10 V and applied to the tip.

**Results and discussions:**

*Epitaxy and morphology of ScAlN nanowires.* Schematically shown in Figure 1a, conventional III-nitrides have a wurtzite (wz) phase ground state, whereas ScN has a rock-salt (rs) phase ground state. This incompatibility leads to phase transition, lattice distortion, and material quality degeneration with increasing Sc content. [16-17] To date, the effect of those evolution trends on domain structure and morphology have remained elusive, because all previous studies were focused on ScAlN films with limited tunability. [17, 45] For nanowire growth, the crystal sites are separate, allowing each crystal phase to nucleate and grow via their most preferable crystallographic phase, thus providing a unique crystalline platform to visualize the transition process. Figure 1b shows the cross-sectional scanning electron microscope (SEM) images of ScAlN/GaN nanowires spontaneously grown on Si(111). The corresponding bird's view SEM images can be found in Supporting Information Figure S1. The Sc content was determined using EDS equipped in the SEM system. By modulating the Sc/Al flux ratio, the Sc content was successfully tuned in the entire compositional range from 0 to 1 (Supporting Information Figure S2). A clear evolution trend of the nanowire morphology is observed with increasing Sc contents: (i) For a Sc content below 0.35, the ScAlN nanowires have a smooth sidewall, similar to conventional AlGaN nanowires, as shown in Figure 1b2-b7 [46]; (ii) For a moderate Sc content (0.4 - 0.5), ScAlN nanocrystals (with a size below 50 nm) start to grow on the sidewall, and the nanocrystal density increases with Sc content, while the nanowires are still dominated by vertical growth, as shown in Figure 1b8-b9; (iii) For Sc contents beyond 0.5, the nanowire growth is dominated by tilted nano-domains (with a size of about 100 nm), and these nano-domains exhibit a clear cubic shape with further increasing Sc content (Figure 1b10-b11). For the ScN/GaN nanowires, regular cubic domains are directly grown on top of the GaN nanowires with a crystal size of ~50 nm. The body diagonal of the ScN cubic domains is parallel with the growth direction, *i.e.*, c-axis (<0001>) of the GaN nanowires, which is determined by the epitaxial relationship between wz-GaN and rs-ScN, *i.e.*, $(0001)[11\bar{2}0]GaN \| (111)[1\bar{1}0]ScN$. [47] As the nanowires can be seen as strain-free for the regions far from the ScAlN/GaN interface, the clear morphology and geometry transition observed in ScAlN nanowires captured the intrinsic phase transition phenomena with increasing Sc content. Figure 1c presents the X-ray diffraction (XRD) $2\theta/\omega$ scans of the nanowire samples shown in Figure 1b. All spectra have been corrected using the diffraction peak of the Si(111) plane. The slight variation of the GaN (0002) diffraction peaks is mainly due to the slight tilting of self-organized GaN nanowires. A clear characteristic diffraction peak in a range of 35.8 – 36.5° is observed for ScAlN nanowires with Sc contents up to 0.4, suggesting a wurtzite crystal structure for those nanowires. However, the wurtzite phase diffraction peak degrades significantly when the Sc content increased to 0.45. Further increasing the Sc content to 0.5, a weak diffraction peak corresponding to the wurtzite phase ScAlN with lower Sc content can be detected, suggesting that a phase separation might have taken place in some regions of the nanowires. For even higher Sc contents ($\geq 0.65$), rs-phase dominated the growth, resulting in a diffraction peak close to the GaN(0002) peak, consistent with previous reports. [48] These phenomena suggest that the phase transition from wurtzite to rock-salt crystal structure happened in a Sc content range of 0.45 - 0.65 depending on the growth conditions and layer thickness, while the wurtzite crystal structure can be well maintained below this compositional range. The gradual transition, however, indicates that the transition is intrinsic of the ternary alloy, consistent with theoretical predictions [16]. Considering that only the wurtzite portion contributes to the piezoelectric response and ferroelectricity of the alloy, our results suggest that the overall piezoelectricity and ferroelectricity can further be tuned in a novel way by controlling the crystal phase portions in the material.

*Microstructure and phases of ScAlN Nanowires.* The crystal structure and phase transition of ScAlN nanowires have been further investigated utilizing STEM. While the microstructure of ScAlN with lower Sc contents (less than 0.4) has been extensively investigated, [49-50] we focus on ScAlN nanostructures with higher Sc contents in this study, where a crystallographic transition from the wurtzite to the cubic is anticipated. Two representative ScAlN nanowire samples with a Sc content of about 0.40 (Figure 1b8) and 0.45 (Figure 1b9) were characterized, shown in Figures 2 and 3, respectively. For all STEM measurements, ScAlN nanowires were mechanically removed from the Si substrate and dispersed on a lacey carbon film mesh Cu TEM grid.

Figure 2a illustrates the high-angle annular dark-field STEM (HAADF-STEM) image of a single ScAlN nanowire with a Sc content of ~0.4 measured by EDS, showing a GaN nanowire template at the bottom with a 200-nm-thick ScAlN segment on top. Due to the limited diffusion ability of Sc and Al adatoms compared to Ga, the diameter for ScAlN increased compared to that of GaN. In addition, similar to spontaneously grown AlN/GaN nanowires by MBE, a thin (less than 10 nm) ScAlN shell formed along the GaN nanowires. [51] Under optimized growth conditions, MBE-grown spontaneous III-nitride nanostructures are free of dislocations, as evidenced by the highly uniform contrast for the bottom GaN segment. However, multi-domain structure was observed in the ScAlN segment, which could be due to the low adatom mobility of Sc and Al. To further identify the crystal phase of the structure, selective area electron diffraction (SAED) patterns were collected from GaN, ScAlN/GaN interface, and top ScAlN regions, which are shown in Figure 2b1-b3, respectively. A single set of bright diffraction spots were observed from GaN segments (b1), suggesting a well-defined wurtzite structure, which was confirmed by comparing the spacing ratio of the (0002) and $(11\bar{2}0)$ plane diffraction spots to the corresponding lattice spacing ratio of ideal wurtzite GaN. The SAED pattern collected from the ScAlN/GaN interface is shown in Figure 2b2. Similar regular diffraction patterns to GaN were observed, indicating that the wurtzite structure was carried over to ScAlN. By examining the image more carefully, two sets of diffraction patterns can be identified, with almost the same pattern spacing in the horizontal direction, suggesting that the ScAlN layer was coherently grown on GaN near the interface region. As a contrast, the top ScAlN segment shows tilted and slightly extended diffraction spots (Figure 2b3), indicating non-uniform domain structure and degraded material quality, which is consistent with the columnar structure exhibited in HAADF-STEM image (Figure 2d).

Figure 2c presents the high resolution HAADF-STEM captured at the ScAlN/GaN interface from the same nanowire. It is seen that ScAlN was not only grown on top of the flat GaN $(000\bar{1})$ surface but also on the $(10\bar{1}0)$ side facets forming a core/shell structure. The geometry of the core/shell structure was visualized by collecting the energy dispersive spectroscopy (EDS) elemental maps along the nanowire (Supporting Information Figure S3). The underlying GaN nanowire template shows a perfect ABABAB wurtzite stacking sequence. Even though the wurtzite stacking sequence was well maintained in most ScAlN regions, an ABCABC stacking sequence was observed near the interface, suggesting the existence of zinc blende (zb) phase ScAlN clusters (enclosed with dashed yellow circles). Since the cubic ScAlN is not the thermodynamically ground state for this Sc content (about 0.4), it may be caused by the localized accumulation of Sc adatoms during growth, which could be supported by the dense nanocrystals on the sidewall of the ScAlN nanowires. We point out that the presented STEM images are projected images of the whole atomic stacks along the zone axis, *i.e.,* along the radial direction, thus the nanocrystals grown on the side wall of ScAlN also contributed to the observed atomic stacking sequence as well as the SAED patterns. In such a case, the observed cubic clusters are likely to come from the nanocrystals grown on the sidewall rather than a cubic domain located inside the ScAlN segment. The STEM image recorded from the ScAlN nanowire near the top surface but away from the sidewall is shown in Figure 2d. Indeed, no obvious ABCABC stacking was found (Supporting Information Figure S4). Instead, clear and sharp c-axis aligned clusters were observed, with a lateral size less than 10 nm. The fact that it is difficult to obtain a clear atomic image for all columns using the same zone axis suggests that these domains have minor in-plane rotation, agreeing well with the SAED pattern shown in Figure 2b3. This kind of multi-domain structure was not observed at the ScAlN/GaN interface. The gradual relaxation of the compressive strain accumulated at the ScAlN/GaN interface could be responsible for the generation of those nano-domains.

The multi-domain structure and the nanocrystal feature became more significant with increasing Sc content to 0.45, as shown in Figure 3a. Meanwhile, elongated diffraction spots are observed in the SAED pattern collected near the interface region (Figure 3b2), suggesting a significant distortion of the wurtzite crystal structure. In the top ScAlN region, the wurtzite-like diffraction pattern was completely substituted by circularly arranged diffraction spots, indicative of misaligned lattice without any preferred orientation. Figure 3c displays the STEM image taken at the ScAlN/GaN interface. Similarly, a few cubic ScAlN domains enclosed with dashed yellow circles appeared, which could again come from the contribution of the sidewall nanocrystals (shown in Figure 1b9). Surprisingly, on the left side of the image, well-ordered wurtzite stacking of ScAlN was clearly identified in a range of at least 10 nm away from the ScAlN/GaN interface (see Supporting Information Figure S5 for more information), manifesting an attractive methodology to prepare high-Sc-content, wurtzite phase ScAlN thin films, even though c-axis aligned lattice was completely absent in the top ScAlN region. Most recently, ferroelectric polarization switching has been demonstrated in nanometer thick wurtzite phase ScAlN films with lower Sc contents (less than 0.3), however, the switching voltages are still high and the margin between the coercive field and breakdown field is still challenging for wurtzite ferroelectrics [39-41]. In theory, the coercive field of ferroelectric ScAlN decreases with increasing Sc content under the wurtzite crystal structure framework. [20] As such, the confined wurtzite phase with high Sc content unveiled in this work represents a pivotal step toward the low voltage operation, thickness scaling down, and nanosized switching of wurtzite ferroelectrics. Our results also address the divergence between the maximum Sc contents for wurtzite phase ScAlN reported by different groups, as the strain state of the layers vary based on the deposition conditions and substrates used. [17, 22, 52]

***Ferroelectric characterization.*** To explore the ferroelectricity of the ScAlN nanowires, displacement current measurements and piezoresponse force microscopy (PFM) were used to analyze the polarization switching characteristics of an exemplary ScAlN nanowire sample with a nominal Sc content of ~0.33. The length of the NWs was extended to have a slightly merged surface to reduce the sidewall leakage after electrode formation. $SiO_2$ deposited by PECVD was used to define the area of the electrode and Ti/Al metal stacks were used as contacts, all of which after poling were further removed by HF for PFM measurements. The SEM images of the sample showing the vertical morphology of the NWs and slightly merged surfaces can be found in the supplementary files (Supporting Information Figure S6). Figure 4a displays the displacement currents measured for the ScAlN NW sample at room temperature on an electrode of 20 μm in diameter. Triangular voltage profiles were used and the leakage current was directly removed using the PUND procedure. [53] With incremental poling voltages, increasing displacement currents were detected, corresponding to more polarization switched regions. The integral of the displacement current, which reveals the polarization that was switched, can further be found in Figure 4b. Due to the non-uniformity of and possibly more leakage pathways between the nanowires, no clear saturation was observed as shown in Figure 4c. The leakage issue may be addressed in the future through careful passivation of the nanowire surfaces. PFM measurements were further conducted to monitor and verify the polarization switching characteristics. Before PFM measurements, the capacitors were poled first following which the top electrode and $SiO_2$ layer were removed using HF. Figures 4e-f delineate the released surface morphology, amplitude contrast, and phase distribution measured by PFM after different voltage poling sequences. The white dashed curve indicates the edge between the electrode and pristine regions (which were protected by $SiO_2$ during poling). Interestingly, ScAlN nanowires poled by -135 V voltage pulses showed reduced amplitude contrast and reversed phase contrast, in comparison to the almost unchanged piezoelectric response for +150 V poled regions. Scattered points instead of uniform contrast in the negatively poled regions are likely a reflection of the non-uniform coercive field, either due to the length or surface variations of the nanowires. In other words, during the poling, only part of the nanowires was switched before PFM measurements. Nevertheless, the obvious difference between the poled regions and the 180° phase separation clearly suggests that the polarity of the nanowires has been switched, at least partially, during the electrical poling. Those results imply that the ScAlN NWs are indeed ferroelectric, rendering them the first wurtzite phase ferroelectric nanowires.

While ferroelectricity in wurtzite phase nitride films have been explored extensively recently, the demonstration of ferroelectric ScAlN nanowires has its unique importance [20]. First, ferroelectrics functioning at reduced dimensionality have long been presumed to construct highly scaled, energy-efficient electronics. [11] The switchable domain size limit in wurtzite ferroelectrics is yet unknown.

Figure 4g suggests, for the first time, that individual domains in the diameter of tens of nanometers can be switched and stabilized in wurtzite ferroelectrics. Second, for some nanowires, the whole ScAlN have been switched, as indicated by the red circles in Figure 4e-f. Considering the intrinsically fast strain relaxation property of the nanowires, and the minimal energy interactions between adjacent nanowires, the results imply that it is possible to intrinsically scale down the size of switchable ScAlN domains down to tens of nanometers, which is of extreme importance for device applications. We note that in some recent publications either the gate length or the distance between the source and drain have been scaled to less than a hundred nanometers. [27, 43] However, the actual size of the switchable domains was not examined. Here, we present the first unambiguous evidence for extreme lateral scaling down of wurtzite ferroelectrics, opening the gate for a wealth of compact, power-efficient electronic and piezoelectric devices.

In summary, we have presented the first comprehensive investigation of the crystallographic phase transition of ScAlN nanowires across the full Sc compositional range. While a gradual transition from the wurtzite to the cubic phase was observed with increasing Sc composition, we further demonstrated that a highly ordered wurtzite phase can be well-confined at the ScAlN/GaN interface, even with Sc content surpassing what can be achieved in thick films. We provide the first distinct evidence of ferroelectric switching in nanometer sized ScAlN nanowires, a result that holds significant implications for future device miniaturization. The successful fabrication of tunable ferroelectric ScAlN nanowires opens new possibilities for nanoscale, domain, alloy, and quantum engineering of ferroelectrics, representing a significant stride towards the development of next-generation, miniaturized devices based on wurtzite ferroelectrics.

**Supporting Information:** Please see the supporting information for the additional images about the microstructure of the ScAlN nanowires.

**Acknowledgement:** This work was supported by National Science Foundation (Award #: 2235377). We are very grateful for Dr. Zhaoying Chen and Boyu Wang for the XRD measurements.

**Figure 1**

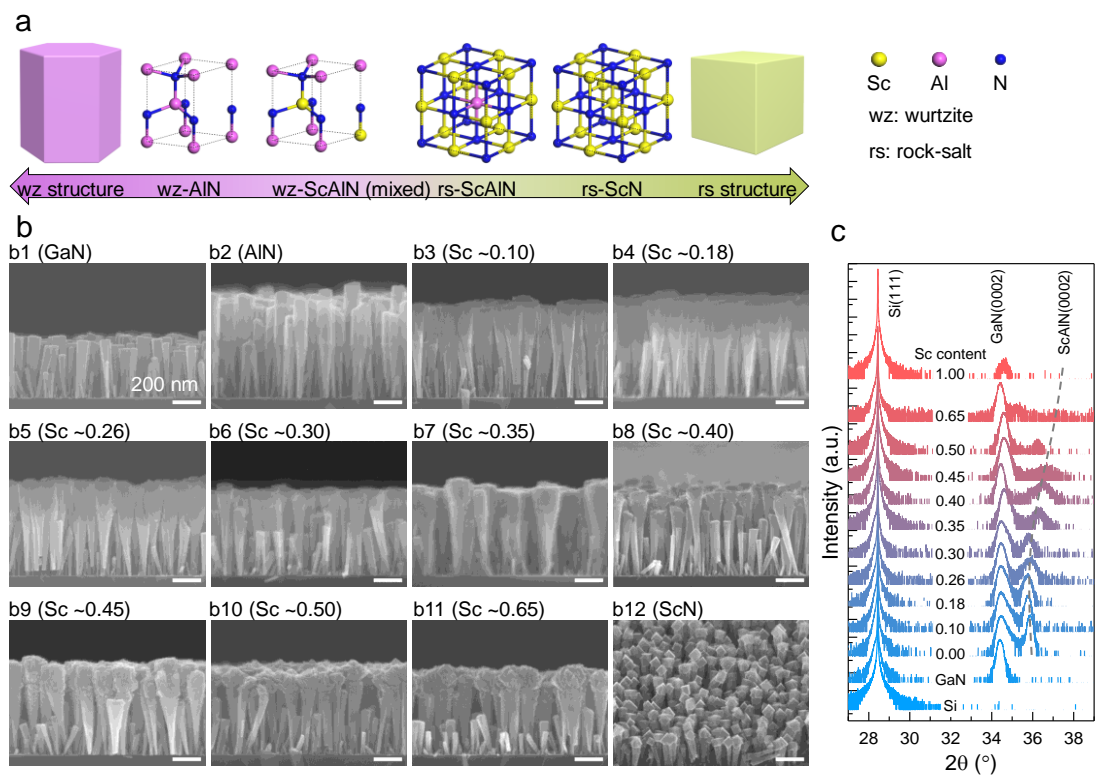

**Figure 1. Structural phase evolution from wurtzite to rock-salt (cubic) in ScAlN nanowires.** (a) Schematic illustration of the geometry and atomic structure of wurtzite and rock-salt phase ScAlN with varying Sc contents. (b) Cross-sectional SEM images of ScAlN nanowire ensemble grown on GaN nanowire templates on Si(111) substrates. The SEM image for ScN/GaN nanowires is 45° tilted to show the unique cubic structure intuitively. (c) XRD 2θ-ω scans for the nanowire samples shown in (b), showing gradually shift towards higher angle side for wurtzite phase ScAlN, while the diffraction peak shift towards the opposite direction for ScAlN nanowires with Sc contents beyond 0.45, due the dominated rock-salt phase. The diffraction peak for Si at 28.4° is used for calibration.

**Figure 2**

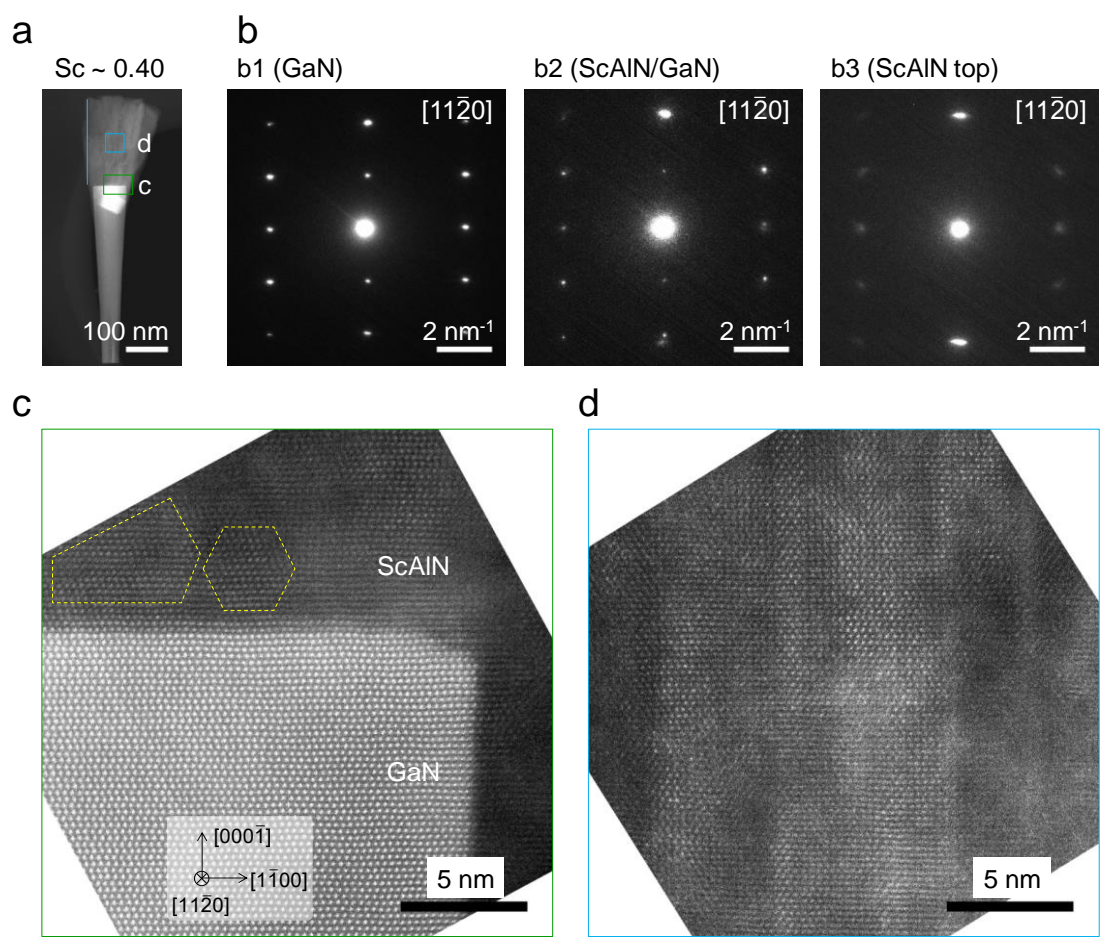

**Figure 2. Structural evolution of wurtzite ScAlN nanowires with a critical Sc content of 0.4.** (a) HAADF-STEM image of a single ScAlN/GaN nanowire, showing GaN at the bottom and ScAlN grown on top with expanded diameter. (b) SAED patterns collected with a zone axis of $[11\bar{2}0]$ from the GaN, ScAlN/GaN, and top ScAlN regions, respectively, confirming the dominated wurtzite structure for the entire nanowire. (c-d) High-resolution HAADF-STEM images recorded from (c) ScAlN/GaN interface and (d) center of ScAlN near the top surface. Some nanometer sized cubic phase ScAlN domains are enclosed with yellow dashed lines in (c). The top ScAlN region exhibits a multi-domain structure with sharp vertical domain boundaries without compromising the wurtzite structure.

**Figure 3**

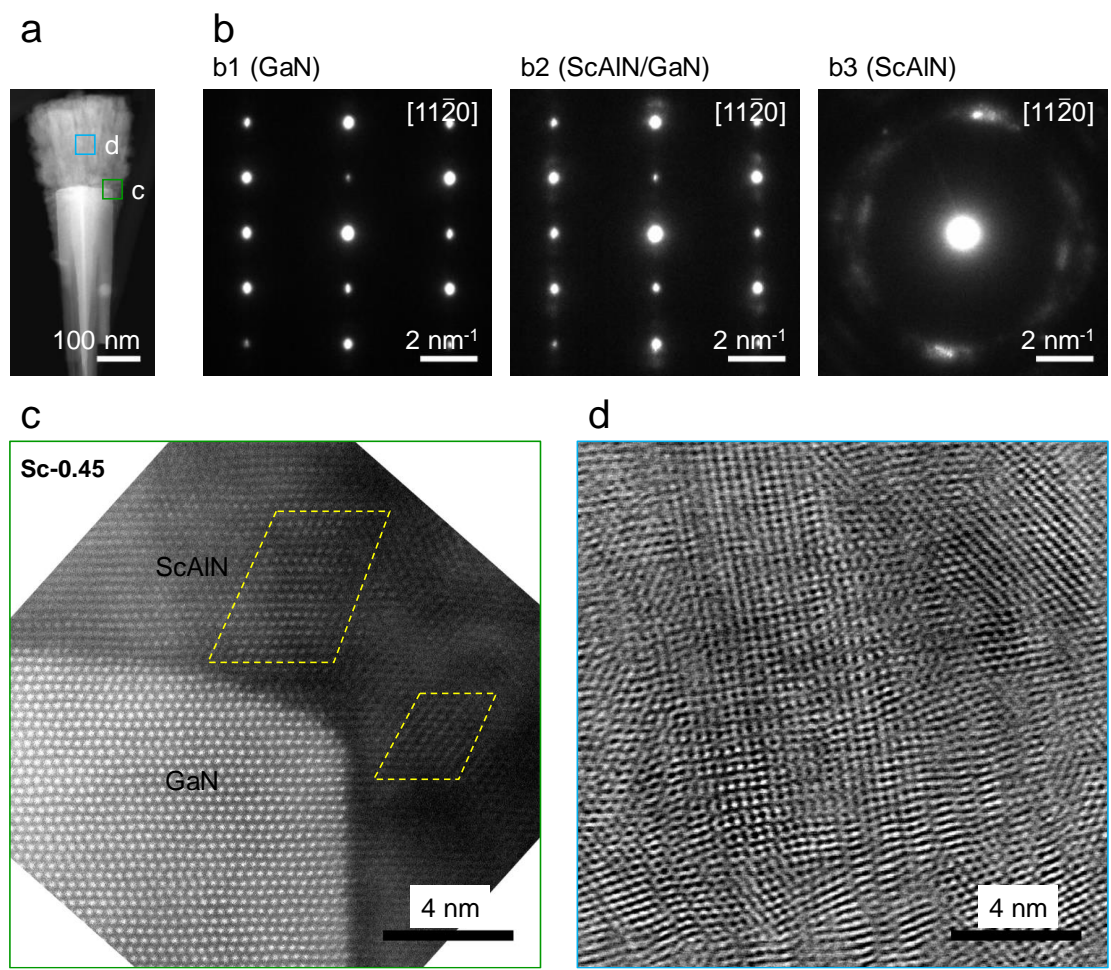

**Figure 3. Structural evolution of wurtzite ScAlN nanowires with a Sc content of 0.5.** (a) HAADF-STEM image of a single ScAlN/GaN nanowire. (b) SAED patterns collected with a zone axis of [11$\bar{2}$0] from the GaN, ScAlN/GaN, and top ScAlN regions, respectively, exhibiting a dominated wurtzite structure at the interface region, while a polycrystalline structure for ScAlN far away from the interface. (c,d) High-resolution HAADF-STEM images recorded from (c) ScAlN/GaN interface and (d) ScAlN near the top surface. Some nanometer sized cubic phase ScAlN domains are enclosed with yellow dashed lines in (c). Wurtzite and cubic mixed phases with varied orientation coexist in the top ScAlN region, giving rise to polycrystalline structure.

**Figure 4**

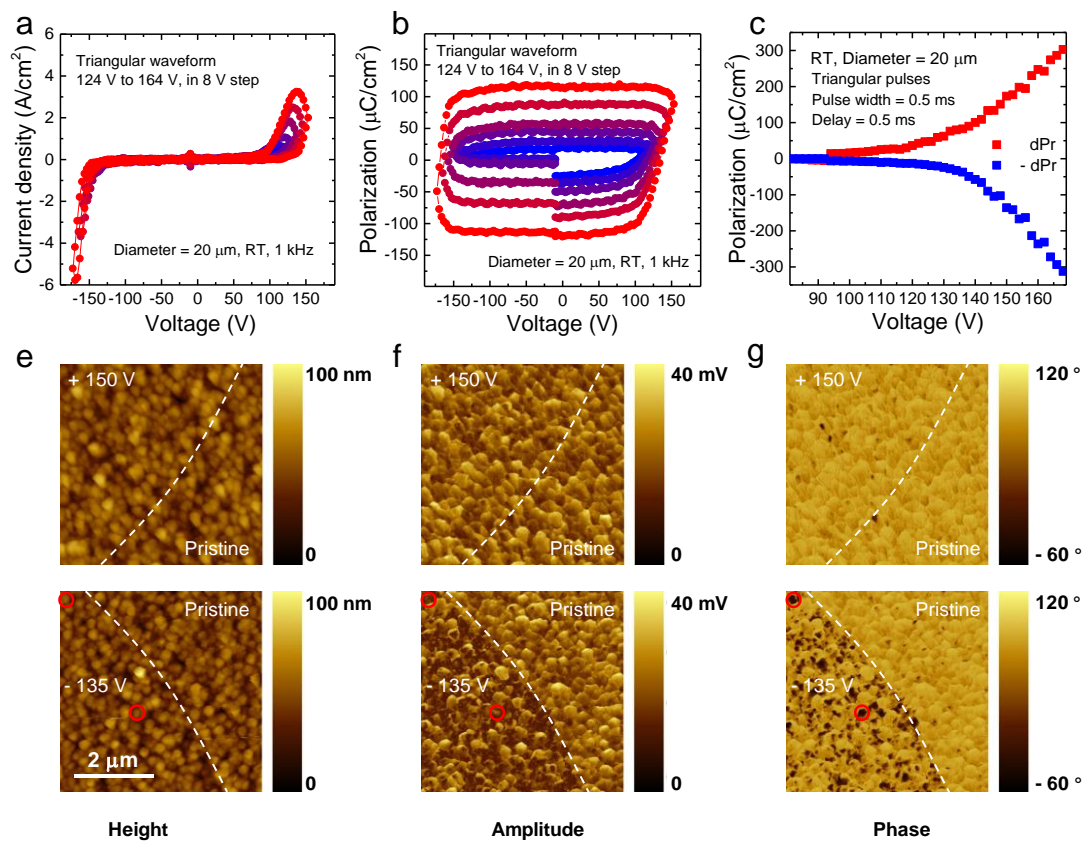

**Figure 4. Ferroelectric properties of the ScAlN nanowires with a Sc content of ~ 0.33.** a) Displacement currents and b) corresponding P-E loops after subtracting non-switching current using a PUND method. c) Voltage-dependent PUND measurement results showing inclining polarization with increasing poling voltages. e) Surface morphology, b) amplitude, and c) phase contrast of the ScAlN nanowires after poling at +150 V (upper panel) and -135 V (lower panel) and removing the top electrodes. The white dashed curves are guidelines for the boundary of the electrode. Red circles indicate switched single nanowires.

**SUPPORTING INFORMATION**

**1. Bird's eye view of SEM images of the nanowires with different Sc compositions**

Due to the low mobility of Sc and Al adatoms, the nanowire diameters increase with Sc and Al incorporation, reducing the gap between nanowires. The nanowire size and spacing is also influenced by the nucleation density of GaN nanowires, depending on the growth conditions.

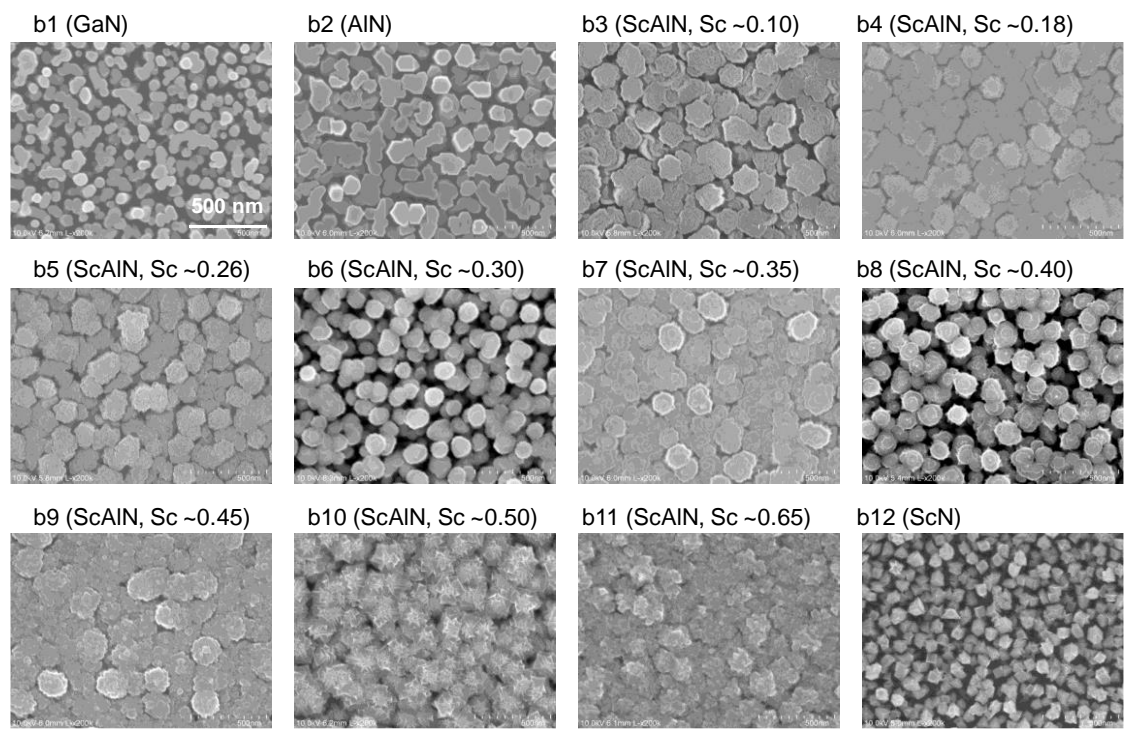

Figure S1. The corresponding top-view SEM images of ScAlN nanowire ensemble grown on GaN nanowire templates on Si (111) substrates (Fig. 1b).

## 2. EDS spectra for ScAlN/GaN nanowires with different Sc contents

EDS was used to determine the Sc composition and the spectra are shown in Figure S2, indicating an increasing Sc signal with increasing Sc composition.

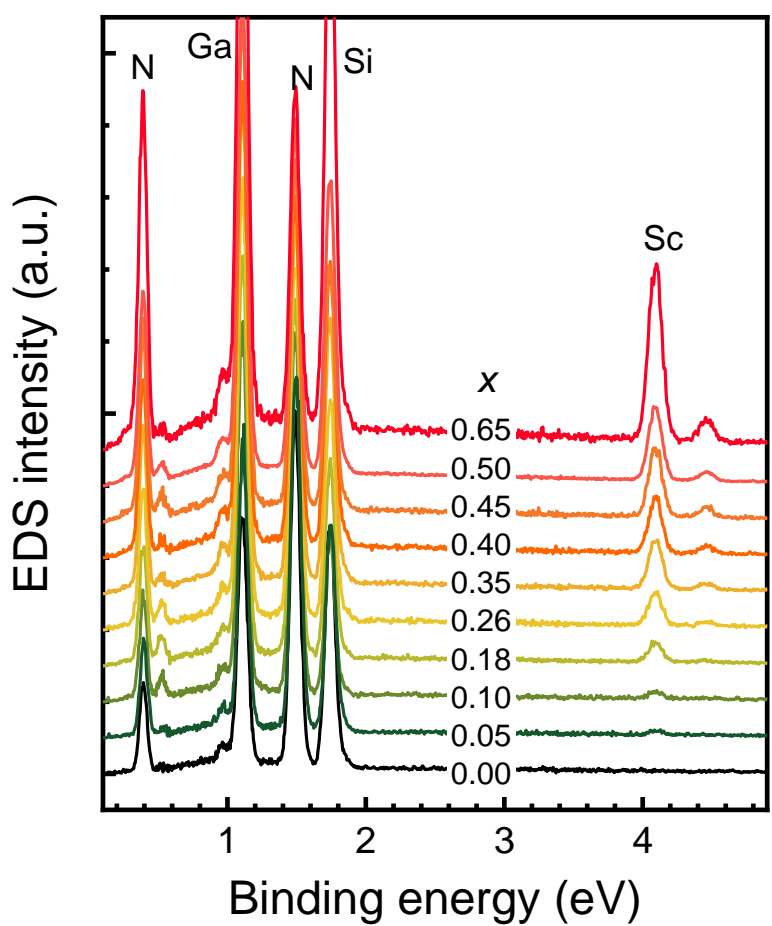

Figure S2. EDS spectra for ScAlN/GaN nanowires with various Sc contents grown on Si(111) substrates.

## 3. Element maps for ScAlN/GaN nanowires

The EDS mapping in TEM for one exemplary ScAlN/GaN nanowire with a nominal Sc content of ~0.4. The low mobility of Sc and Al adatoms results in the formation of a thin ScAlN shell surrounding the GaN nanowires. The high oxygen concentration is due to the easy oxidation of Sc and Al.

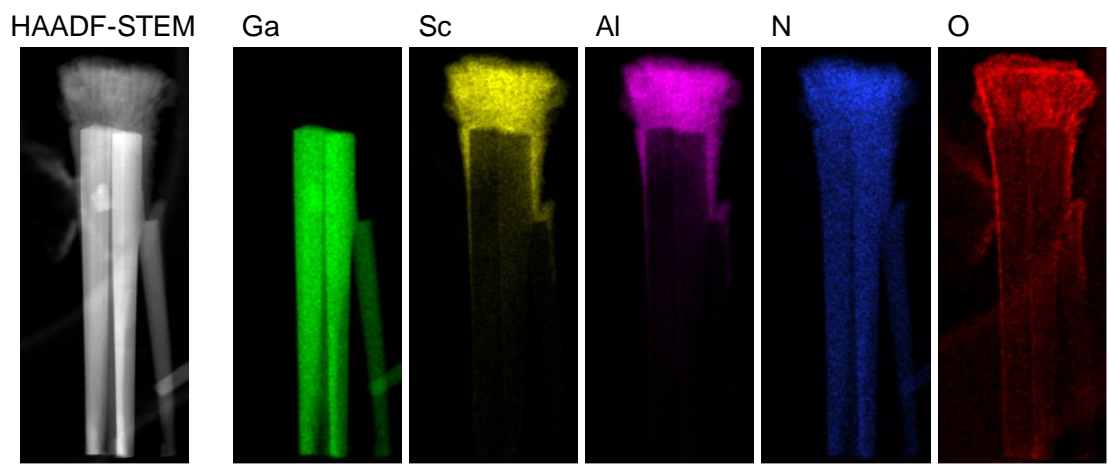

Figure S3. EDS element maps for ScAlN/GaN nanowires, showing a clear core-shell geometry, as well as a clear surface oxidation layer.

## 4. FFT images of the TEM image shown in Figure 2d.

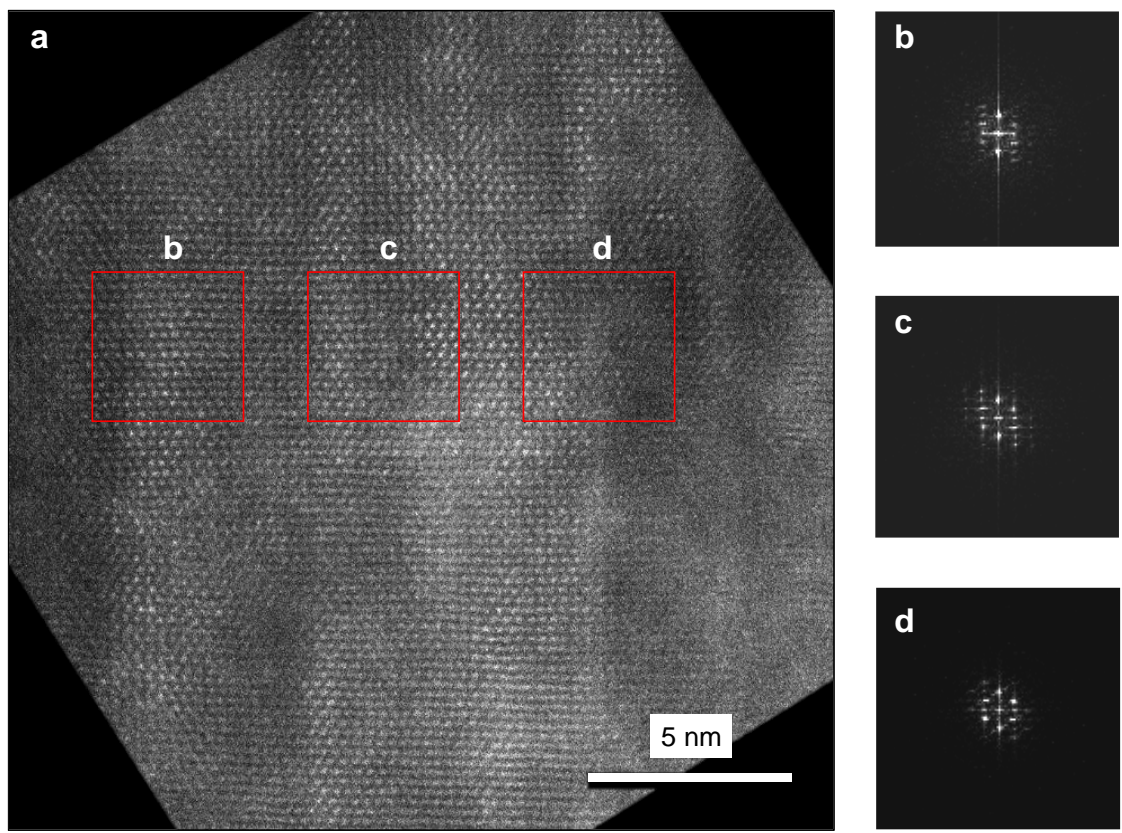

Figure S4. FFT images of different regions shown in Figure 2d, indicating wurtzite structure in the core region of the ScAlN nanowire with a Sc content of ~ 0.4.

5. FFT images of different regions shown in Figure 3c

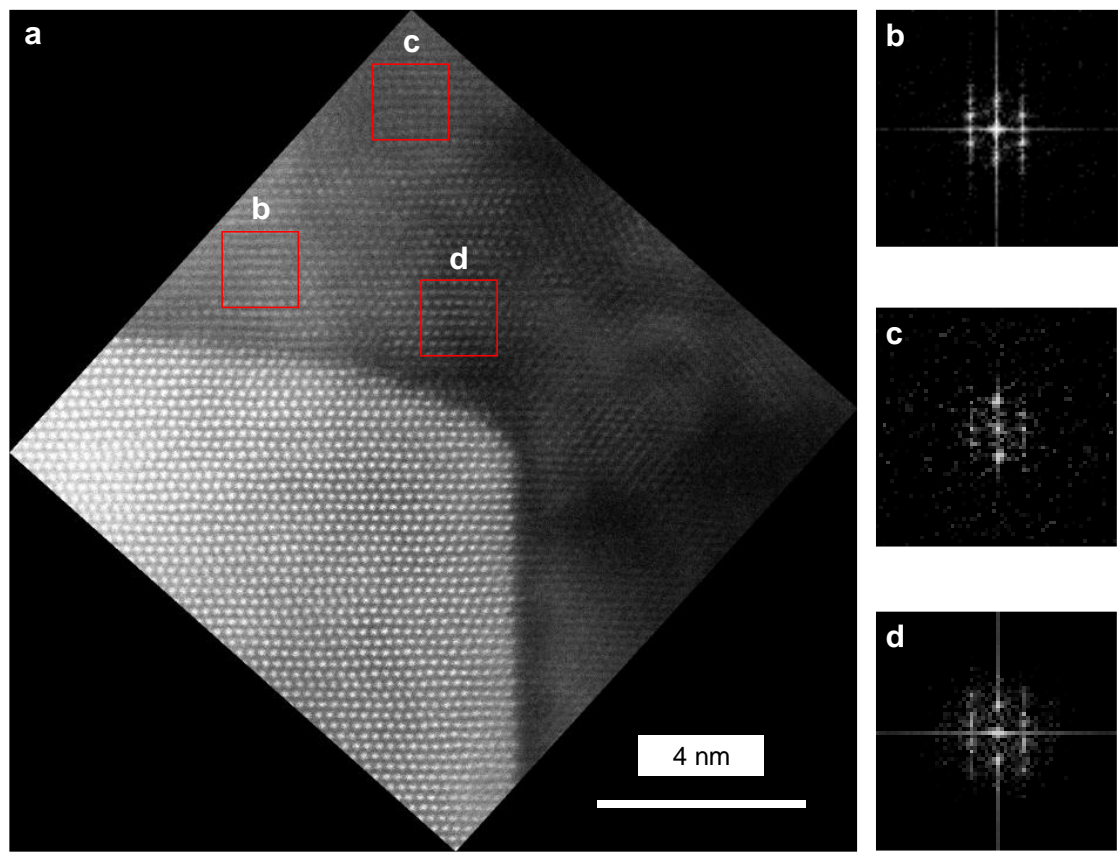

Figure S5. FFT images of different regions shown in Figure 3c, indicating wurtzite ScAlN confined at the interface of ScAlN/GaN with a Sc content of ~ 0.45.

## 6. Microscopic morphology of the ScAlN/GaN nanowires for ferroelectric measurements

To reduce the leakage, relatively thicker ScAlN (~ 300 nm) was grown to minimize the gap between nanowires, and the metal deposition was done by placing the sample perpendicular to the metal flux.

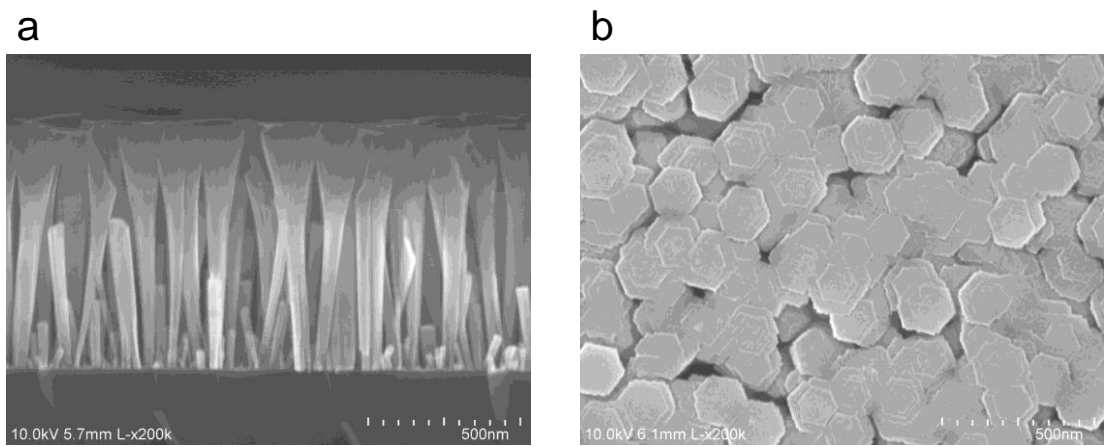

Figure S6. (a) Cross-sectional and (b) top-view SEM images of ScAlN/GaN nanowires grown on Si that were employed for electrical measurements, showing slightly merged surface.